\documentclass{article}
\usepackage[utf8]{inputenc}
\usepackage[T1]{fontenc}
\usepackage[bookmarks,hidelinks]{hyperref}
\usepackage{bookmark}
\usepackage[british]{babel}
\usepackage{comment}
\usepackage{authblk,cite}

\usepackage[legalpaper, margin=3cm]{geometry}

\usepackage{amsmath,amsthm,amsfonts,relsize,bm} %maths stuff
\usepackage{etoolbox}
\usepackage[shortlabels]{enumitem}
\usepackage{xcolor}
\usepackage{centernot}
\theoremstyle{definition}
\newtheorem{definition}{Definition}[section]

\theoremstyle{theorem}
\newtheorem{theorem}{Theorem}[section]

\numberwithin{equation}{section} % to number equations within sections
\newcommand{\ed}{\mathrm{d}}
\newcommand{\ChristSymb}[2]{\big\{^{\hspace{0.2em}#1}_{#2}\big\}} % Christoffel Symbols

\newcommand{\Riem}[3][R]{{#1}^{#2}_{\hspace{0.5em}#3}}

\newcommand{\Lag}{\mathcal{L}} % Lagrangian density
\newcommand{\tr}[1]{\mathrm{tr}\left\{#1\right\}} % trace
\newcommand{\sqrtg}{\sqrt{\vert g\vert}}

\newcommand{\Lie}[1]{\mathlarger{\pounds}_{#1}}

\makeatletter
\newcommand{\subalign}[1]{%
	\vcenter{%
		\Let@ \restore@math@cr \default@tag
		\baselineskip\fontdimen10 \scriptfont\tw@
		\advance\baselineskip\fontdimen12 \scriptfont\tw@
		\lineskip\thr@@\fontdimen8 \scriptfont\thr@@
		\lineskiplimit\lineskip
		\ialign{\hfil$\m@th\scriptstyle##$&$\m@th\scriptstyle{}##$\hfil\crcr
			#1\crcr
		}%
	}%
}
\makeatother

\begin{document}

\title{An Ostrogradsky Instability Analysis of Non-minimally Coupled Weyl Connection Gravity Theories}

\author[1]{Rodrigo Baptista\thanks{\href{mailto:rodrigo.baptista@fc.up.pt}{rodrigo.baptista@fc.up.pt}}\textsuperscript{,}}
\affil[1]{Departamento de Física e Astronomia, Faculdade de Ciências da Universidade do Porto, Rua do Campo Alegre 687, 4169-007 , Porto, Portugal}

\author[1,2]{Orfeu Bertolami\thanks{\href{mailto:orfeu.bertolami@fc.up.pt}{orfeu.bertolami@fc.up.pt}}\textsuperscript{,}}
\affil[2]{Centro de Física das Universidades do Minho e do Porto, Rua do Campo Alegre 687, 4169-007 , Porto, Portugal}

\date{9$^\text{th}$ of September, 2020}

\maketitle
\vspace*{-0.5cm}
\begin{abstract}
	We study the Hamiltonian formalism of the non-minimally coupled Weyl connection gravity (NMCWCG) in order to check whether Ostrogradsky instabilities are present.
	The Hamiltonian of the NMCWCG theories is obtained by foliating space-time into a real line (representing time) and \mbox{3-dimensional} space-like hypersurfaces, and by considering the spatial metric and the extrinsic curvature of the hypersurfaces as the canonical coordinates of the theory.
	Given the fact that the theory we study contains an additional dynamical vector field compared to the usual NMC models, which do not have Ostrogradsky instabilities, we are able to construct an effective theory without these instabilities, by constraining this Weyl field.
	\vspace{1cm}
\end{abstract}

\vspace*{-.5cm}

\section{Introduction}

\hspace*{\parindent}%
Non-minimally coupled curvature-matter gravity theories (NMC)~\cite{Bertolami-NMC} are a natural extension of General Relativity,
where the linear term $R$ is generalised to an arbitrary function, $f_1(R)$, and where a second function $f_2(R)$ is introduced and coupled to the Lagrangian of the matter fields.
Thus, the action of the NMC model is
\begin{equation}\label{NMC-action}
	S=\int_M\Big(\kappa f_1(R) +f_2(R)\Lag\Big)\sqrtg\,\ed^4x,
\end{equation}
where $\kappa=\frac{c^4}{16\pi G}$ and $g$ is the determinant of the metric.
This theory has interesting features, as it admits inflationary solutions in cosmology~\cite{Gomes:2016cwj}, mimicking dark energy and dark matter~\cite{Bertolami-2010,Bertolami-2012,Bertolami-2011.1,Bertolami-Paramos} (see Ref.~\cite{Bertolami:2013xda} for a review).
Further interesting properties arise when adding a non-compatible connection,\linebreak in particular a Weyl connection, to this NMC model.
The resulting theory, dubbed as non-minimally coupled Weyl connection gravity (NMCWCG) has been previously studied in Refs.~\cite{Gomes-Bertolami,Baptista-Bertolami}.\linebreak
This paper follows the work done in previous references and focuses on the Hamiltonian of the NMCWCG theories and on searching for potential Ostrogradsky instabilities.

On his 1850's work, Mikhail Ostrogradsky~\cite{Ostrogradsky} detailed a particular type of instabilities arising on theories described by a Lagrangian depending on higher order derivatives with respect to time.
Given that we expect energy-bounded theories, this type of instabilities constrain the field theories since many of them (and, in particular, many gravitation theories) include higher order derivatives.
Nevertheless, many of these theories are free from Ostrogradsky instabilities by violating one of the assumptions Ostrogradsky's theorem.
Hence, we seek whether the introduction of a Weyl connection with an associated vector field and its dynamics into the NMC model (a theory that does not contain an Ostrogradsky instability) gives rise to this type of instabilities and under which conditions can these instabilities be avoided.

The paper is structured as follows: Sections 2 and 3 review the Non-minimally coupled Weyl connection gravity theory, following the ideas of Refs.~\cite{Gomes-Bertolami,Baptista-Bertolami}, and present Ostrogradsky's theorem for high-order theories, as discussed in Refs.~\cite{Ostrogradsky,Woodard,Querella};
Section 4 constructs a space-time foliation for theories with a non-compatible connection following procedure done for the compatible case \cite{Querella,Gron-Hervik};
Finally, in Section 5, we analyse Ostrogradsky instabilities in the NMCWCG theory.
We draw our conclusions in Section 6.

%------------------------------------------------------------------------------------------
\newpage
\section{Non-minimally Coupled Weyl Connection Gravity}

\hspace*{\parindent}%
A Weyl connection introduces, through the covariant derivative $D_\mu$, a vector field $A_\mu$ such that it is non-compatible with the metric:
\begin{equation}\label{def:WeylG-coord}
	D_\lambda g_{\mu\nu}=A_\lambda\,g_{\mu\nu}.
\end{equation}
This covariant derivative can be written as:
\begin{equation}\label{WeylConnection-LCConnection}
	D_\lambda g_{\mu\nu}=\partial_\lambda g_{\mu\nu}-\bar{\Gamma}^\sigma_{\lambda\mu}g_{\sigma\nu}-\bar{\Gamma}^\sigma_{\lambda\nu}g_{\mu\sigma},
\end{equation}
with connection coefficients given by:
\begin{equation}\label{WeylConnCoeff}
	\bar{\Gamma}^\lambda_{\mu\nu}=\ChristSymb{\lambda}{\mu\nu}-\frac{1}{2}g^{\lambda\sigma}(A_\mu g_{\nu\sigma}+A_\nu g_{\mu\sigma}-A_\sigma g_{\mu\nu}),
\end{equation}
where $\ChristSymb{\lambda}{\mu\nu}$ are the Christoffel symbols.

We assume, for a matter of generality, that the Weyl vector is a non-abelian field with a field strength given by:
\begin{equation}\label{def:Weyl-StrengthTensor}
	F_{\mu\nu}=\partial_\mu A_\nu-\partial_\nu A_\mu-[A_\mu,A_\nu],
\end{equation}
where $[\cdot,\cdot]$ represents the commutator.
Naturally, if $A_\mu$ is considered to be an abelian field, the second term in the previous expression vanishes.
Notice that the generalisation of the Weyl connection to be non-abelian is particularly relevant for admiring the resulting gravity theory in the context of homogeneous and isotropic cosmologies (see Refs.~\cite{Bento:1992wy,Bertolami:1991,Bertolami:1991ff} for extensive discussions). 

The Lagrangian density of the vector field is given, as usual, by:
\begin{equation}\label{def:YM-Lagrangian}
	\Lag_{\rm W}[A_\mu,g^{\mu\nu}] =-\frac{1}{4\mu}\tr{F_{\alpha\beta}F^{\alpha\beta}}-V[A],
\end{equation}
where $\mu$ is equivalent to the electromagnetic permeability, and a potential is admitted.

From Eqs.~\eqref{WeylConnection-LCConnection} and \eqref{WeylConnCoeff}, the Riemann tensor can be obtained.
Contracting this tensor, we get the Ricci tensor:
\begin{equation}
	\bar{R}_{\mu\nu}=R_{\mu\nu}+\frac{1}{2}g_{\mu\nu}\left(\nabla_\lambda-A_\lambda\right)A^\lambda-\frac{3}{2}A_\mu A_\nu-F_{\mu\nu}+\frac{1}{2}E_{\mu\nu},
\end{equation}
with $R_{\mu\nu}$ the Ricci tensor for the Levi-Civita connection;
$\nabla_\mu$ the usual Levi-Civita covariant derivative, $E_{\mu\nu}=\nabla_\mu A_\nu+\nabla_\nu A_\mu+2\left\{A_\mu,A_\nu\right\}$ and $\{\cdot,\cdot\}$ representing the anti-commutator.
Finally, the scalar curvature is given by:
\begin{equation}\label{Scal-Curv-Weyl}
	\bar{R}=R+3\nabla^\lambda A_\lambda-\frac{3}{2}A^\lambda A_\lambda,
\end{equation}
where $R$ is the scalar curvature corresponding to the Levi-Civita connection.

The non-minimally coupled Weyl connection gravity theory (NMCWCG) arises by introducing the Weyl connection into the the non-minimally coupled curvature-matter gravity (NMC):
\begin{equation}\label{Action-NMCWCG}
	S=\int_M\left[\kappa f_1(\bar{R}) -f_2(\bar{R})\left(\frac{1}{4\mu}\tr{F_{\alpha\beta}F^{\alpha\beta}}+V[A]\right)\right]\sqrtg\,\ed^4x.
\end{equation}
Notice that when merging the NMC action, Eq.~\eqref{NMC-action}, and the Weyl connection, the scalar curvature is the one given by Eq.~\eqref{Scal-Curv-Weyl}.
It is this action whose potential Ostrogradsky instabilities we shall analyse.

%------------------------------------------------------------------------------------------
\section{Ostrogradsky's Model and Problem}\label{Sec:Ostro-problem}

\hspace*{\parindent}%
We start by introducing some basic concepts regarding the structure of the system in order to perform an Ostrogradsky analysis\footnote{In this paper we adopt the standard symplectic form for the space $(Q,P)$.}:
\begin{definition}
	A Lagrangian of the type $L(q_i,\dot{q}_i,\ldots,q_i^{(\alpha_i)})$ is called a \textit{high-order Lagrangian} if $\alpha_i>1$ for any $i=1,\ldots,N$, where $q_i^{(k)}$ represent the $k$-th time derivative of $q_i$;
	A Lagrangian $L(q_i,\dot{q}_i,\ldots,q_i^{(\alpha_i)})$ such that $\frac{\partial^2 L} {\partial\big(q_i^{(\alpha_i)}\big)^2}\neq 0$ is said to be \textit{non-degenerate}.
\end{definition}

\noindent{}Hence, Ostrogradsky's theorem of instabilities follow:
\begin{theorem}\label{Theorem:Ostro}
	Let $L(q_i,\dot{q}_i,\ldots,q_i^{(\alpha_i)})$ be a non-degenerate high-order Lagrangian of a given physical system. Then, the system has unbounded states of energy.
\end{theorem}
\noindent{}The proof of this statement, which follows the ideas of Ref.~\cite{Woodard}, starts by reducing the Lagrangian of this system to an equivalent one, but only in first order in the derivatives of the coordinates.
Thus, allowing the phase space coordinates to be
\begin{subequations}
\begin{align}
	Q_i^{k_i} &= q_i^{(k_i-1)},\\[0.5em]
	P_i^{k_i} = \frac{\partial L}{\partial\big(q_i^{(k_i)}\big)}+&\sum_{p_i=1}^{\alpha_i-k_i}\left(-\frac{\ed}{\ed t}\right)^{p_i}\frac{\partial L}{\partial\big(q_i^{(p_i+k_i)}\big)},
\end{align}
\end{subequations}
for all $i$ and $k_i=1,\ldots,\alpha_i-1$, we can compute the Hamiltonian for the system.
Taking into account that by hypothesis, the Lagrangian is non-degenerate, the equation $P_i^{\alpha_i}=\frac{\partial L}{\partial\big(q_i^{(\alpha_i)}\big)}$ has the solution
\begin{equation}
	\dot{Q}_i^{\alpha_i}=q_i^{(\alpha_i)} =\chi_i(Q_j^1,\ldots,Q_j^{\alpha_j},P_j^{\alpha_j}).
\end{equation}
Thus, the Hamiltonian is given by:
\begin{equation}
	H(Q,P)=\sum_{i=1}^{N}\left(\sum_{k_i=1}^{\alpha_i-1}P_i^{k_i}\,\dot{Q}_i^{k_i}+P_i^{\alpha_i}\chi_i\right)-L(Q_i^1,\ldots,Q_i^{\alpha_i},\chi_i),
\end{equation}
where we see that the dependence of $H$ on the generalised momenta, $P_i^1$, $\ldots$, $P_i^{\alpha_i-1}$, appears only linearly.
This means that, because this Hamiltonian is a constant of motion\footnote{This comes from the hypothesis that the Lagrangian does not dependent explicitly on time.}, we can choose a configuration of the system that has $P_i^1$, for example, as small as we want and, naturally, the energy as negative as we want.

This way to prove Theorem~\ref{Theorem:Ostro}, follows Ostrogradsky's own proposal for canonical coordinates.
However, there is a more general way to prove this theorem if we instead assume the following coordinate transformation:
\begin{equation}\label{Ostr-canon-coord}
	Q_i^0=q_i,\quad Q_i^{\beta_i}=\xi_i^{\beta_i}(q_j,\dot{q}_j,\ldots,q_j^{(\theta_{ij})}),\;\theta_{ij}=\min\{\beta_i,\alpha_j-1\},
\end{equation}
with $\xi_i^{\beta_i}$ invertible.
Then, by following the ideas of Refs.~\cite{Querella,Henneaux-Teitelboim,Oksanen}, the canonical Hamiltonian is given by:
\begin{equation}\label{Hamil-constr}
	H_c(Q,P)=\sum_{i=1}^N\left(\sum_{\beta_i=0}^{\alpha-2} P_i^{\beta_i}\mathcal{Q}_i^{\beta_i}+P_i^\alpha \dot{Q}_i^\alpha\right)-L(Q,\dot{Q}),
\end{equation}
with, 
\begin{equation}
	P_j^{\beta_j}=\frac{\partial}{\partial (\dot{Q}_j^{\beta_j})}\left(L(Q_i^{\beta_i},\dot{Q}_i^\alpha)+\sum_{j=1}^N\sum_{\beta_j=0}^{\alpha-1}(\dot{Q}_j^{\beta_j-1}-\mathcal{Q}_j^{\beta_j})\lambda_j^{\beta_j}\right),
\end{equation}
where $\mathcal{Q}_i^{\beta_i}(Q_j^0,\ldots,Q_j^{\beta_j})=\dot{Q}_i^{\beta_i}$ is obtained by inverting Eq.~\eqref{Ostr-canon-coord} and reintroducing it in the derivative of Eq.~\eqref{Ostr-canon-coord} where $\lambda_j^{\beta_j}$ are Lagrange multipliers.
Once again, Eq.~\eqref{Hamil-constr} contains Ostrogradsky instabilities (the second term inside parenthesis).
This second procedure more closely resembles the ideas we shall use when searching for Ostrogradsky instabilities in the NMCWCG model.

Notice that theories described by a degenerate Lagrangian might still contain Ostrogradsky instabilities due to the lower order (yet greater than first order) derivatives \cite{Motohashi-1,Motohashi-2,Motohashi-3,Motohashi-4}.
Our analysis is sufficiently general to address these cases as well.

%------------------------------------------------------------------------------------------
\section{A Space-time Foliation}\label{Sec:3+1Space-time_Split}

\hspace*{\parindent}%
A fundamental step to check whether the NMCWCG model contains Ostrogradsky instabilities is the foliation\footnote{This foliation is necessary since we require the Hamiltonian of the NMCWCG theory and this formalism depends on singling out time (which we shall denote as $\mathbf t$).}
of space-time into a real line guided by a time-like 4-vector $\mathbf{t}=t^\mu\,\partial_\mu$ and a family of spatial hypersurfaces $\Sigma_t$, indexed by time, with spatial metric $h_{\mu\nu}$ and normal (unit) 4-vector $\mathbf{n}=n^\mu\,\partial_\mu$.
This means that only NMCWCG field configurations compatible with the chosen space-\linebreak-time symmetries will be considered and all degrees will either follow the relevant field equations or a set of constraints.
The present section follows closely Chapter~14 of Ref.~\cite{Gron-Hervik} as well as Section~3.3 of Ref.~\cite{Querella}.
Both references treated the case $\mathcal{D}_\lambda g_{\mu\nu}=0$\footnote{In this section we use the notation $\mathcal D$ for a generic covariant derivative. Only in the next section we shall recall it as the Weyl connection.
For this reason, we shall call $\Riem[\hat{R}]{\lambda}{\mu\nu\sigma}$ the Riemann tensor corresponding to the connection $\mathcal D$.}.
We, instead, consider the more general case: $\mathcal{D}_\lambda g_{\mu\nu}\neq0$.

Firstly, we may relate $t^\mu$ and $n^\mu$ as
\begin{equation}
	t^\mu=Nn^\mu+N^\mu,
\end{equation}
where $N$ and $\mathbf{N}=N^\mu\,\partial_\mu$ are the lapse function and shift vector, respectively.
The metric tensor can be expressed as:
\begin{equation}\label{OstrInst:Metrics}
	g_{\mu\nu}=h_{\mu\nu}-n_\mu n_\nu,
\end{equation}

Notice that we can compute different quantities in two different bases for the manifold, $M=\mathbb{R}\times\Sigma_t$:
\begin{enumerate}[(i)]
	\item The standard basis $\{\partial_\mu\,\vert\,\mu=0,\ldots,3\}$;
	\item A basis $\{\mathbf{e}_{\tilde a}\,\vert\,\tilde{a}=1,\ldots,3\}$ of $\Sigma_t$ together with the vector $\mathbf n$.
\end{enumerate}
For convenience, elements of the item (ii) are always denoted with a tilde.
Thus, basis (ii) can be written in a compact form as:
\begin{equation}
	\{\mathbf{e}_{\tilde\mu}\,\vert\,\tilde{\mu}=0,\ldots,3\}=\{\mathbf{e}_{\tilde a}\,\vert\,\tilde{a}=1,2,3\}\cup\{\mathbf{n}\},\quad{\rm and}\quad\mathbf{e}_{\tilde 0}\equiv\mathbf{n}.
\end{equation}
The definitions we have introduced so far allows us to define the derivative with respect to time as the Lie derivative in the direction of the vector $\mathbf{t}$:
\begin{equation}\label{Timeder}
	\frac{\ed}{\ed t}\equiv \Lie{\mathbf t}.
\end{equation}
Furthermore, we can define an extrinsic curvature of $\Sigma_t$, with components\footnote{In Eq.~\eqref{def:ExtrinsicCurv} we have used the metric as the linear map $\mathbf{g}(p):T_pM\times T_pM\to\mathbb{R},\ \forall\,p\in M$, where $T_pM$ is the tangent space of the space-time manifold $M$ at a point $p\in M$. In the following sections, we use this definition more often.}:
\begin{equation}\label{def:ExtrinsicCurv}
	K_{\tilde{a}\tilde{b}}=-h_{\tilde a}^\mu h_{\tilde b}^\nu\,\mathbf{g}\left(\mathbf{e}_{\mu},\mathcal{D}_\nu\mathbf{n}\right).
\end{equation}
We can simplify Eq.~\eqref{def:ExtrinsicCurv} in two ways: first, by using the definition of the basis (ii); and by computing the covariant derivative of the metric $\mathcal{D}_{\tilde b}\mathbf{g}(\vec{e}_{\tilde a},\vec{n})$.
These yield:
\begin{equation}\label{ExtrinsicCurv-2ways}
	K_{\tilde{a}\tilde{b}}=-\Gamma_{\tilde{b}\tilde{a}\tilde{0}}=(\mathcal{D}_{\tilde b}\mathbf{g})(\mathbf{e}_{\tilde a},\mathbf{n})-\Gamma^{\tilde 0}_{\tilde{b}\tilde{a}}.
\end{equation}

In order to complete this description, we must express the scalar curvature of $M$ written in terms of the scalar curvature of the spatial hypersurfaces $\Sigma_t$, $\tilde{R}$, the extrinsic curvature, $K_{\tilde{a}\tilde{b}}$ and its derivatives as well as derivatives of the spatial metric $h_{\tilde{a}\tilde{b}}$.

Using the definition of the Riemann tensor with respect to the connection coefficients\footnote{
	In this definition we have used that, given two tensors $T_\mu$ and $S_\mu$, $T_{[\mu}S_{\nu]}\equiv\frac{1}{2}(T_\mu S_\nu-T_\nu S_\mu)$.
	In a later expression we shall use the similar definition $T_{(\mu}S_{\nu)}\equiv\frac{1}{2}(T_\mu S_\nu+T_\nu S_\mu)$.
},
\begin{equation}
	\Riem[\hat{R}]{\lambda}{\mu\sigma\nu}=2\partial_{[\sigma}\Gamma^\lambda_{\nu]\mu}+2\Gamma^\lambda_{\alpha[\sigma}\Gamma^\alpha_{\nu]\mu},
\end{equation}
and by embedding it in $\Sigma_t$, we obtain
\begin{equation}\label{Riemann-Embedded}
	\Riem[\hat{R}]{\tilde a}{\tilde{b}\tilde{c}\tilde{d}}=\Riem[\tilde{R}]{\tilde a}{\tilde{b}\tilde{c}\tilde{d}}+2\Gamma^{\tilde a}_{\tilde{0}[\tilde{c}}\Gamma^{\tilde 0}_{\tilde{d}]\tilde{b}}.
\end{equation}
Turning Eq.~\eqref{Riemann-Embedded} into an equivalent covariant expression and contracting to obtain the scalar curvature, we get
\begin{align}\label{OstrInst:ScalCurv-Incomp}
	\hat{R}&=\tilde{R}+K^2-K^{\alpha\beta}K_{\alpha\beta}-2\hat{R}_{\alpha\beta}n^\alpha n^\beta\nonumber\\
	&\hspace{2em}+K g^{\beta\delta} (\mathcal{D}_{\beta}\mathbf{g})(\mathbf{e}_{\delta},\mathbf{n})-K^{\alpha\beta}(\mathcal{D}_{\beta}\mathbf{g})(\mathbf{e}_{\alpha},\mathbf{n}),
\end{align}
where $\hat{R}_{\alpha\beta}$ is the Ricci tensor.
After simplifying the term $\hat{R}_{\alpha\beta}n^\alpha n^\beta$, Eq.~\eqref{OstrInst:ScalCurv-Incomp} becomes
\begin{align}\label{OstrInst:ScalarCurvature}
	\hat{R}&=\tilde{R}+K^2-3K^{\alpha\beta}K_{\alpha\beta}-2h^{\tilde{a}\tilde{b}}\Lie{\mathbf n}K_{\tilde{a}\tilde{b}}-2(\mathcal{D}_{\mathbf n}h^{\tilde{a}\tilde{b}})K_{\tilde{a}\tilde{b}}-K(\mathcal{D}_{\mathbf n}\mathbf{g})(\mathbf{n},\mathbf{n})\nonumber\\
	&\hspace{3em}-Kh^{\beta\delta}(\mathcal{D}_\beta\mathbf{g})(\mathbf{e}_\delta,\mathbf{n})+K^{\alpha\beta}(\mathcal{D}_\beta\mathbf{g})(\mathbf{e}_\alpha,\mathbf{n}),
\end{align}
where $\mathcal{D}_{\mathbf n}\equiv n^\mu\,\mathcal{D}_\mu$.

Finally, by using properties of the Lie derivative and definition \eqref{OstrInst:Metrics}, we find that
\begin{equation}\label{Lieder-Metrich}
	\big(\Lie{\mathbf t}\mathbf{h}\big)(\partial_\mu,\partial_\nu) =(\mathcal{D}_{\mathbf t}\mathbf{h})(\partial_\mu,\partial_\nu) -2NK_{(\mu\nu)}+2h_{\alpha(\mu}\mathcal{D}_{\nu)}N^\alpha.
\end{equation}

%------------------------------------------------------------------------------------------
\newpage
\section{Ostrogradsky Instabilities in the NMCWCG theory}

\hspace*{\parindent}%
When studying the NMCWCG model, we must first choose the canonical coordinates.
Taking into account the study of the previous section, the natural choices for the canonical coordinates are
\begin{equation}\label{Canon-Coord}
	(q_0)_{\tilde{a}\tilde{b}}=h_{\tilde{a}\tilde{b}},\quad
	(q_1)_{\tilde{a}\tilde{b}}=K_{\tilde{a}\tilde{b}},
\end{equation}
constrained to Eq.~\eqref{Lieder-Metrich}.
This choice yields a quite standard Hamiltonian formalism from which we can compute the extended Lagrangian by introducing Lagrange multipliers to set the constraint:
\begin{equation}\label{ExtLagrange}
	\Lag^\ast=N\sqrt{h}\,\Lag+\lambda^{\tilde{a}\tilde{b}}\left(\Lie{\mathbf t}(q_0)_{\tilde{a}\tilde{b}} +2N(q_1)_{\tilde{a}\tilde{b}}-2(q_0)_{\tilde{c}(\tilde{a}}D_{\tilde{b})}N^{\tilde c}\right),
\end{equation}
where $\Lag$ is the Lagrangian of the NMCWCG model, Eq.~\eqref{Action-NMCWCG}, that is the term inside square brackets.
After Eq.~\eqref{ExtLagrange}, we drop $(q_0)_{\tilde{a}\tilde{b}}$ and $(q_1)_{\tilde{a}\tilde{b}}$ in favour of the more intuitive $h_{\tilde{a}\tilde{b}}$ and $K_{\tilde{a}\tilde{b}}$\footnote{Notice that now $\mathcal{D}_\lambda=D_\lambda$ and $\Riem[\hat{R}]{\lambda}{\mu\nu\sigma}=\Riem[\bar{R}]{\lambda}{\mu\nu\sigma}$, etc.}.

Before obtaining the canonical momenta, we notice that the existence of the Weyl connection still admits that $(D_{\tilde b}\mathbf{g})(\mathbf{e}_{\tilde a},\mathbf{n})=A_{\tilde b}\,\mathbf{g}(\mathbf{e}_{\tilde a},\mathbf{n})=0$ since $\mathbf n$ is a normal vector to $\Sigma_t$.
Also, we can compute the covariant derivative of $h^{\mu\nu}$ as
\begin{equation}
	D_{\mathbf n}g^{\mu\nu}=-\mathbf{A}(\mathbf{n})g^{\mu\nu}\implies D_{\mathbf n}h^{\mu\nu}=-\mathbf{A}(\mathbf{n})(h^{\mu\nu}-2n^\mu n^\nu),
\end{equation}
where $\mathbf{A}(\mathbf{n})=A_\alpha n^\alpha=A_ {\tilde 0}$ and, since we only work with spatial components $\tilde{a}\tilde{b}$, the second term inside parenthesis vanishes.
Thus, the scalar curvature $\bar{R}$ is given by:
\begin{equation}\label{ScalCurv-splitted-NMCWCG}
	\bar{R}=\tilde{R}+K^2-3K^{\alpha\beta}K_{\alpha\beta}-2h^{\tilde{a}\tilde{b}}\Lie{\mathbf n}K_{\tilde{a}\tilde{b}}+3K\mathbf{A}(\mathbf{n}).
\end{equation}

Now, we proceed with the computation of the Hamiltonian of the NMCWCG model.
The conjugate momenta to the canonical coordinates in Eq.~\eqref{Canon-Coord} are\footnote{In Eq.~\eqref{canon-momenta} we are using the dot as the time derivative following Eq.~\eqref{Timeder}.}
\begin{equation}\label{canon-momenta}
	p_0^{\tilde{a}\tilde{b}}=\frac{\partial\Lag^\ast}{\partial\dot{h}_{\tilde{a}\tilde{b}}}=\lambda^{\tilde{a}\tilde{b}},\qquad p_1^{\tilde{a}\tilde{b}}=\frac{\partial\Lag^\ast}{\partial\dot{K}_{\tilde{a}\tilde{b}}}=-2\kappa\sqrt{h}h^{\tilde{a}\tilde{b}}\Theta,
\end{equation}
with $\Theta=f_1'(\bar{R})+\kappa f_2'(\bar{R})\,\Lag$, where the prime represents the derivative with respect to $\bar{R}$.
Similarly, we find the canonical momenta for the Weyl vector field,
\begin{equation}
	\Pi^{\mu}=\frac{\partial \Lag^\ast}{\partial\dot{A}_\mu}=N\sqrt{h}\,f_2(\bar{R})\pi^{\mu},
\end{equation}
where $\pi^{\mu}=\frac{\partial}{\partial\dot{A}_\mu}\Lag_{\rm W}$.
The canonical Hamiltonian is then given by:
\begin{equation}\label{CanonHamil-start}
	H_c=\int_{\Sigma_t}\Big[\,p_0^{\tilde{a}\tilde{b}}\dot{h}_{\tilde{a}\tilde{b}}+p_1^{\tilde{a}\tilde{b}}\dot{K}_{\tilde{a}\tilde{b}}+\Pi^{\mu}\dot{A}_\mu-N\sqrt{h}\left(\kappa f_1(\bar{R})+f_2(\bar{R})\Lag_{\rm W}\right)\Big]\,\ed^3x.
\end{equation}

To obtain an explicit expression of the canonical Hamiltonian, we start using the definition of $p_1^{\tilde{a}\tilde{b}}$, Eq.~\eqref{canon-momenta}, and invert Eq.~\eqref{ScalCurv-splitted-NMCWCG}.
Thus Eq.~\eqref{CanonHamil-start} becomes:
\begin{align}\label{CanonHamil-interm}
	H_c&=\int_{\Sigma_t}\Bigg[p_0^{\tilde{a}\tilde{b}}\dot{h}_{\tilde{a}\tilde{b}}+\kappa N\sqrt{h}\,\Theta\left(\bar{R}-\tilde{R}-K^2+3K^{\tilde{a}\tilde{b}}K_{\tilde{a}\tilde{b}}-3K\mathbf{A}(\mathbf{n})-\frac{2h^{\tilde{a}\tilde{b}}}{N}\Lie{\mathbf N}K_{\tilde{a}\tilde{b}}\right)\nonumber\\
	&\hspace{3em}-N\sqrt{h}\left(\kappa f_1(\bar{R})+f_2(\bar{R})\mathcal{E}\right)\Bigg]\,\ed^3x,
\end{align}
where we have defined:
\begin{equation}
	\mathcal{E}\equiv\dot{A}_\mu\frac{\partial\Lag_{\rm W}}{\partial\dot{A}_\mu}-\Lag_{\rm W}.
\end{equation}
Grouping together similar terms in Eq.~\eqref{CanonHamil-interm}, we obtain the canonical Hamiltonian as:
\begin{align}\label{CanonHamil-final}
	H_c&=\int_{\Sigma_t}\Bigg[\kappa N\sqrt{h}\left(\Theta\,\bar{R}-f_1(\bar{R})-\frac{f_2(\bar{R})}{\kappa}\mathcal{E}\right)+\frac{N}{6}p_1(\tilde{R}+K^2-3K_{\tilde{a}\tilde{b}}+3K\mathbf{A}(\mathbf{n}))\nonumber\\
	&\hspace{3em}+\frac{h^{\tilde{a}\tilde{b}}}{3}p_1(\Lie{\mathbf N}K_{\tilde{a}\tilde{b}})+p_0^{\tilde{a}\tilde{b}}\left(D_{\mathbf t}h_{\tilde{a}\tilde{b}}-2NK_{\tilde{a}\tilde{b}}+2h_{\tilde{c}(\tilde{a}}D_{\tilde{b})}N^{\tilde c}\right)\Bigg]\,\ed^3x.
\end{align}

\pagebreak
The conditions for time slicing and the structure of Eq.~\eqref{CanonHamil-final} are similar to the one in General Relativity: $H_{\rm GR}=\int_{\Sigma_t}(N\mathfrak{H}+N^{\tilde a}\mathfrak{H}_{\tilde a})\,\ed^3x$, i.e. it is a sum of constraints.
These two terms are the super-Hamiltonian, $\mathfrak{H}$, and the super-momentum, $\mathfrak{H}_{\tilde a}$.
Since the super-momentum concerns Lorentz invariance, the Ostrogradsky instabilities appear only on the super-Hamiltonian of the theory, thus, we need only to search for the terms that depend on $N$ and not in $\mathbf N$.
The super-Hamiltonian of the NMCWCG model is, therefore,
\begin{align}\label{superHami-NMCWCG}
	\mathfrak{H}&=\kappa\sqrt{h}\left[\Theta(\bar{R})-f_1(\bar{R})-\frac{f_2(\bar{R})}{\kappa}\mathcal{E}\right]%_{\subalign{&\bar{R}=G(p_1)\\&\Pi_m=\frac{\partial\Lag}{\partial(\dot{\Psi}_m)}}}
	+\frac{p_1}{6}(\tilde{R}+K^2-3K_{\tilde{a}\tilde{b}}K^{\tilde{a}\tilde{b}})\nonumber\\
	&\hspace{1em}-2p_0^{\tilde{a}\tilde{b}}K_{\tilde{a}\tilde{b}}+\frac{1}{2}p_1 K\mathbf{A}(\mathbf{n})+p_0^{\tilde{a}\tilde{b}}D_{\mathbf n}h_{\tilde{a}\tilde{b}}.
\end{align}

As seen in Section~\ref{Sec:Ostro-problem}, the Ostrogradsky instability appears as oddly powered terms of momenta in the Hamiltonian.
By solving second class constraints (see Ref.~\cite{Querella} for further details), $p_0^{\tilde{a}\tilde{b}}=\sqrt{h}K^{\tilde{a}\tilde{b}}$ and $p_1^{\tilde{a}\tilde{b}}=-2\sqrt{h}h^{\tilde{a}\tilde{b}}$ which means that, when searching for Ostrogradsky instabilities, it is equivalent to look for odd powered terms in the extrinsic curvature.
Analysing Eq.~\eqref{superHami-NMCWCG}, we find two terms that are linear in the extrinsic curvature/canonical momentum\footnote{Notice that, if we make $A_\mu=0$ in all previous equations of this Section, there are no odd powered terms of the extrinsic curvature and, hence, no Ostrogradsky instability in NMC gravity theories (Eq.~\eqref{NMC-action}), as expected.}: $K\mathbf{A}(\mathbf{n})$ and $p_0^{\tilde{a}\tilde{b}}D_{\mathbf n}h_{\tilde{a}\tilde{b}}$.
These terms give us an expression that constrains the system so to be free from Ostrogradsky instabilities:
\begin{equation}\label{OstrInst:ConstraintNMCWG}
	\frac{1}{2}p_1 K\mathbf{A}(\mathbf{n})+p_0^{\tilde{a}\tilde{b}}D_{\mathbf n}h_{\tilde{a}\tilde{b}}=0.
\end{equation}
By using Eq.~\eqref{def:WeylG-coord},
\begin{equation}
	D_{\mathbf n}h_{\mu\nu}=\mathbf{A}(\mathbf{n})(h_{\mu\nu}+2n_\mu n_\nu),
\end{equation}
which means we can rewrite Eq.~\eqref{OstrInst:ConstraintNMCWG} as:
\begin{equation}
	\left(\frac{1}{2}p_1+\sqrt{h}\right)K\mathbf{A}(\mathbf{n})=0.
\end{equation}
Since the term inside parenthesis does not vanish, because $p_1=-6\sqrt{h}$, and the solution $K=0$ is uninteresting, the only possible solution that discards an Ostrogradsky instability is $\mathbf{A}(\mathbf{n})=A_\alpha n^\alpha=0$. Written in the basis (ii) introduced in Section \ref{Sec:3+1Space-time_Split},
\begin{equation}\label{Ostr:Constraint}
	A_{\tilde 0}=0,
\end{equation}
which constrains the effective theories arising from the NMCWCG model to those where the Weyl vector has only spatial components, $A_{\tilde\mu}=(0,A_{\tilde a})$.

%------------------------------------------------------------------------------------------
\vspace{1cm}
\section{Conclusion}

\hspace*{\parindent}%
In this work, we have examined the issue of Ostrogradsky instabilities in the non-minimally coupled Weyl connection gravity (NMCWCG) by studying its Hamiltonian formulation.
In order to achieve that, we foliated space-time into a real line guided by a time-like vector and 3-dimensional spatial Cauchy surfaces.
The foliation of space-time allows for a definition of a spatial metric and an extrinsic curvature which are the canonical coordinates of the system, constrained to Eq.~\eqref{Lieder-Metrich}, and a suitable Hamiltonian.

For the NMCWCG model, we find that the super-Hamiltonian contains linear terms on the conjugate momenta, which imply that the theory contains Ostrogradsky instabilities.
However, since the system contains an arbitrary vector field, we can constrain it in order to avoid the instabilities arising from the problematic terms in the super-Hamiltonian.
This is accomplished by imposing the condition $A_{\tilde 0}=0$ (the Weyl field has no `time' component).
This is a relevant result since the difference between the theory studied in this work and the one from Ref.~\cite{Gomes-Bertolami} is the existence of dynamics of the Weyl field.
Hence, the introduction of this dynamical term in the Lagrangian of the theory induces a constraint on the field which limits the number of consistent effective theories that are free from Ostrogradsky instabilities.

%------------------------------------------------------------------------------------------

\end{document}